\documentclass[aps,prx,showkeys,amsmath,amssymb,reprint,superscriptaddress,noeprint]{revtex4-2}
\usepackage{graphicx}
\usepackage{xcolor}
\usepackage{amsmath}
\usepackage{amssymb}
\usepackage{braket}
\usepackage{subfigure}
\usepackage[pdftex, colorlinks=true, linkcolor=black, citecolor=black, urlcolor=black]{hyperref}

\newcommand{\orange}[1]{\textcolor{orange}{#1}}

\begin{document}

\title{Quantum Chaos as an Essential Resource for Full Quantum State Controllability}

\newcommand{\RegensburgUniversity}{Institut f\"ur Theoretische Physik, Universit\"at Regensburg, D-93040 Regensburg, Germany}
\newcommand{\Steve}{Department of Physics and Astronomy, Washington State University, Pullman, WA USA}

\author{Lukas Beringer}
\email{lukas.beringer@physik.uni-regensburg.de}
\affiliation{\RegensburgUniversity}
\author{Mathias Steinhuber}
\affiliation{\RegensburgUniversity}
\author{Klaus Richter}
\affiliation{\RegensburgUniversity}
\author{Steven Tomsovic}
\affiliation{\RegensburgUniversity}
\affiliation{\Steve}

\begin{abstract}

Using the key properties of chaos, i.e.~ergodicity and exponential instability, as a resource to control classical dynamics has a long and considerable history. However, in the context of controlling ``chaotic'' quantum unitary dynamics, the situation is far more tenuous. The classical concepts of exponential sensitivity to trajectory initial conditions and ergodicity do not directly translate into quantum unitary evolution. Nevertheless properties inherent to quantum chaos can take on those roles:  i) the dynamical sensitivity to weak perturbations, measured by the fidelity decay, serves a similar purpose as the classical sensitivity to initial conditions;  and ii) paired with the fact that quantum chaotic systems are conjectured to be statistically described by random matrix theory, implies a method to translate the ergodic feature into the control of quantum dynamics. With those two properties, it can be argued that quantum chaotic dynamical systems, in principle, allow for full controllability beyond a characteristic time that scales only logarithmically with system size and $\hbar^{-1}$. In the spirit of classical targeting, it implies that it is possible to fine tune the immense quantum interference with weak perturbations and steer the system from any initial state into any desired target state, subject to constraints imposed by conserved quantities. In contrast, integrable dynamics possess neither ergodicity nor exponential instability, and thus the weak perturbations apparently must break the integrability for control purposes. The main ideas are illustrated with the quantum kicked rotor. The production of revivals, cat-like entangled states, and the transition from any random state to any other random state is possible as demonstrated.  

\end{abstract}


%%%%%%%%%%%%%%%%%%%%%%%%%%%%%%%%%%%%%%%%%%%%%%%%%%%%%%%%%%%%%%%%%%%%
\maketitle

\section{Introduction}
\label{sec:intro}

Out-of-equilibrium classically chaotic dynamical systems are naturally expected to relax, equilibrate, or thermalize rapidly.  They evolve ergodically, approaching arbitrarily close to every available phase space point, and any initially localized continuous density of initial conditions spreads exponentially rapidly, eventually covering the available phase space uniformly.  Guiding such systems toward some desired precise final state would seem to pose insurmountable barriers.  However, it has long been shown how classical chaos can be considered a resource for efficient control.  More precisely, in the theory of {\it classical targeting}~\cite{Ulam58, Ott90, Shinbrot90, Kostelich93, Bollt95, Schroer97} ergodicity and exponential instability guarantee fast transport pathways between any two available states, requiring only small perturbations. 

The unitary evolution of quantum mechanics is distinctly different from classical dynamical evolution.  Despite various attempts at defining quantum Lyapunov exponents or dynamical entropy~\cite{Connes87, Alicki94, Lindblad88}, the unitary dynamics of bounded, isolated quantum systems is not exponentially unstable, i.e.~effectively Lyapunov exponents must vanish, and unitary evolution is reversible in a way that classical dynamics is not~\cite{Shepelyansky83, Tomsovic16}.  Furthermore, the quantum concept of ergodicity cannot rely on the existence of a phase space, and depending upon the context, ergodicity in quantum systems can take on slightly different meanings; more on this below.

%%%%%%%%%%%%%%%%%%%%%%%%%%%%%%%%%%%%%%%%%%%%%%%%%%%%%%%%%%%%%%%%%%%%%
\begin{figure}
    \centering
    \includegraphics[width=0.75\linewidth]{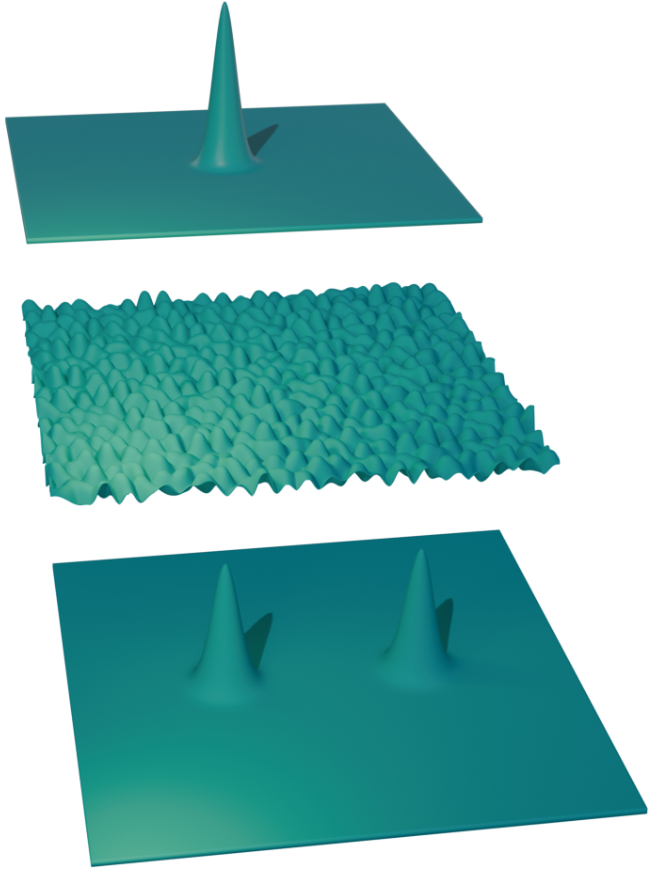}
    \caption{Schematic illustration of quantum chaos as a resource for control.  In a quantum chaotic system, an initially localized state (upper panel),  after a very brief evolution time, relaxes into a complicated state (center) that   is distributed over the available energy surface.  In the same way, however, chaos enables a weak perturbation of the system such that shortly afterward the state converges to any desired final state, for example a cat-like entangled state (lower panel).}
    \label{fig:schematic}
\end{figure}
%%%%%%%%%%%%%%%%%%%%%%%%%%%%%%%%%%%%%%%%%%%%%%%%%%%%%%%%%%%%%%%%%%%%%%

The absence of strict quantum equivalents of classical ergodicity and exponential sensitivity renders the concept of chaos in quantum mechanics as somewhat problematic.  Nevertheless, the useful moniker, {\it quantum chaos} (itself an oxymoron), transmits the important notion that there are manifestations of chaos in quantum mechanics for systems with classical chaotic limits and special properties even for those without.  It is therefore natural to ask whether quantum chaos can serve an analogous purpose for controlling quantum unitary dynamics as ergodicity and instability do for controlling chaotic classical dynamics.

In the rapidly advancing field of quantum technologies and quantum computation, the ability to steer the dynamics of quantum systems plays a crucial role. Prominent methods, such as quantum optimal control~\cite{Tannor85, Brumer86, Peirce88, Werschnik07, Doria11, Boscain21, Ansel24, Koch22} and shortcuts to adiabaticity~\cite{Demirplak03, Berry09, Delcampo17, Gueryodelin19}, enable the high precision control of quantum states through tailored time-dependent external fields.  Compared to classical dynamics though, little is known about how quantum chaos affects controllability of quantum dynamics; see however~\cite{Gong05, Gruebele07, Poggi16}. Recently it was shown that the concepts of classical targeting can be extended to guide quantum states along fast solutions of their classical limit~\cite{Tomsovic23, Tomsovic23b, Beringer24}. Although effectively an interesting and potentially useful coherent approach to quantum simulation, so far it is restricted to Schrödinger quantum dynamics and many-body bosonic systems with rather localized quantum states which are constrained to remain so. This naturally excludes many interesting possibilities such as entangled states or many-body fermionic and spin systems.

The purpose of this work is to demonstrate that quantum chaos does provide a resource for control even without an existing classical limit, relying on analogous ergodicity and exponential instability properties applicable in a general quantum setting. More specifically the universal behavior of quantum chaotic dynamics enables the fine tuning of the huge amount of quantum interference inherent to their unitary evolution, in order to rapidly control the system into arbitrary target quantum states. A precursor to this approach is the identification of quantum revivals in a fully chaotic quantum stadium billiard~\cite{Tomsovic97}.

Thus, quantum chaos as a resource conceptually would signify that whatever the initial state of interest, it swiftly devolves into some kind of a random-like state, only to reform seemingly magically and rapidly to the desired final state.  This is schematically illustrated in Fig.~\ref{fig:schematic} in which a localized state disperses rapidly into a complicated random-like state only to reemerge in a very short time scale as a localized cat-like entangled state.  The evolving state spreads wildly with many parts interfering with other parts, but by design small perturbations generate the necessary interferences leading to the desired target state.  The main challenge is making small alterations to all the quantum interferences in such a way as to arrive at the desired target state. 

This paper is structured as follows. Section \ref{sec:fidrmt} examines the translation of the classical concepts of ergodicity and instability into quantum mechanics.  In Sec.~\ref{sec:kr} quantum chaos as a resource for full state controllability is illustrated with the quantum kicked rotor. In particular, the ideas are demonstrated for a revival, as well as the preparation of a cat-like and a completely random state.  Section \ref{sec:sum} ends with a summary and possible directions of future research.

%%%%%%%%%%%%%%%%%%%%%%%%%%%%%%%%%%%%%%%%%%%%%%%%%%%%%%%%%%%%%%%%%%%%%%

\section{Instability and Ergodicity in Quantum Systems}
\label{sec:fidrmt}

As prerequisites for the application to quantum control, two generic properties of quantum chaotic systems are introduced. First, the sensitivity to weak perturbations as measured by the fidelity decay implies relevant scales for the weakness of the control parameters and the swiftness of the control time.  Combining that  with an ergodic property enabling a construction of dynamical ensembles, which mimic essential properties found in random matrix ensembles, full state controllability emerges.

\subsection{Dynamical sensitivity to weak perturbations and relevant control times}
\label{sec:instab}

In classical chaotic targeting a key ingredient is the ability to steer the state of a system with only small perturbations, relying on the exponential sensitivity of the dynamics as given by the Kolmogorov-Sinai entropy, $h_{\rm KS}$,~\cite{Kolmogorov58, Sinai59} i.e.~the sum of positive Lyapunov exponents, $\sum_j \lambda_j$~\cite{Pesin77}.  For quantum systems a measure of their sensitivity to perturbations is Peres' fidelity decay~\cite{Peres84b} \begin{equation}
    \label{eq:fidel}
    \mathcal{C}_\alpha(t) = |\braket{\alpha|{\rm e}^{i\widehat{H}t/\hbar}{\rm e}^{-i\widehat{H}_0t/\hbar}|\alpha}|^2 \ ,
\end{equation}
also referred to as the \textit{Loschmidt echo}~\cite{Jalabert01, Goussev12}, with
\begin{equation}
    \label{eq:Hamfid}
    \widehat{H} = \widehat{H}_0 + \epsilon \widehat{H}_c \ ,
\end{equation}
an unperturbed Hamiltonian $\widehat{H}_0$ and an $\epsilon$-dependent weak generic perturbation $\widehat{H}_c$, possibly time-dependent. 
The unperturbed energy expectation value of the initial state $|\alpha\rangle$ is assumed to be $E$. Semiclassical theory applied to quantum chaotic dynamical systems gives precise answers in terms of perturbation strength regimes and rates of decay in terms of $\epsilon$, $h_{\text{KS}}$, and classical action diffusion constants~\cite{Cerruti02, Cerruti03}, with the greatest concern here being in the Fermi Golden rule and Lyapunov regimes~\cite{Jalabert01,Gutkin10}.  In fact, it is actually the $\epsilon$ value at the boundary between these two regimes that, in the case where a semiclassical limit exists, is of greatest interest; see arguments ahead.
 
In terms of quantum control, an arbitrary initial state in a chaotic dynamical system would rapidly evolve into a state that typically has an overlap with the desired target state on the order of $1/N$, where $N$ is the effective dimensionality of the Hilbert space to which the initial and target state belong. Consider the case in which the energy width of the initial and target states correspond approximately to the Thouless energy~\cite{Edwards72}, $\Delta E_{\rm Th}$, then 
\begin{equation}
\label{eq:Nth}
 N \approx \Delta E_{\rm Th} \rho_g(E) \approx {\cal V}_{\rm Th}/g h^D    
\end{equation}
($D$ being the number of degrees of freedom and $g$ being a factor accounting for some symmetry), which is the Thouless energy multiplied by the symmetry specific density of states, $\rho_g(E)$. This is also approximately the total phase space volume, ${\cal V}_{\rm Th}$, within the Thouless energy window divided by a Planck cell size of $h^D$ and $g$. For a generic case, in order for the initial state to be transferred into the target state, the system's dynamics needs to be sufficiently altered by the control time $t^*$ that $C_{\alpha}(t)$ has roughly decayed to $1/N$. In quantum chaotic systems with the perturbation strong enough to be in the Fermi Golden rule regime, the exponential decay rate is given by  $\Gamma=2\pi\rho_g(E)\epsilon^2 v_{\rm rms}^2/\hbar$, where $v_{\text{rms}}$ is the root mean square (rms) of the matrix elements of $\widehat H_c$.

For a given $\epsilon$, the time by which the perturbation has sufficiently altered the dynamics (such that $C_{\alpha}(t)$ has decayed to roughly $1/N$) is then given by
\begin{equation}\label{eq:min_time}
    t = \frac{\hbar \ln N}{2\pi v_{\rm rms}^2\rho_g(E) \epsilon^2}\ .
\end{equation}
In the context of quantum control, this constitutes the minimal time to steer an initial state into any desired target state by weakly perturbing the system. Usual notions of \textit{quantum speed limits} based on the minimum uncertainty relation~\cite{Mandelstam45, Margolus98, Caneva09, Deffner17}, like the Mandelstam-Tamm bound, give the minimal time to move away from itself, whereas Eq.~\eqref{eq:min_time} represents the time scale at which every possible state in a quantum chaotic system can in principle be reached.

In those quantum chaotic systems for which a classical limit exists, a critical time scale is the so-called logtime~\cite{Berman78, Berry79b},
\begin{equation}
\label{eq:logt}
\tau = \frac{1}{h_{\rm KS}} \ln N\ ,
\end{equation}
which can be thought of crudely as the shortest time scale it takes for an initially localized or nearly minimal uncertainty state to spread to all the phase space cells.  It should be noted that this is a very short time scale, which grows very slowly with system size, numbers of degrees of freedom, and $\hbar\to 0$; see Appendix~\ref{sec:app2}.  This has a strong influence on the fidelity decay.  As $\epsilon$ is increased within the Fermi Golden rule regime, the decay gets more and more rapid until it saturates at the Lyapunov regime~\cite{Jalabert01} with the decay governed by $h_{\rm KS}$ and no longer dependent on the strength $\epsilon$.  So again Eq.~\eqref{eq:logt} is the shortest workable control time scale that can be arranged with a vanishingly small perturbation as $\hbar\rightarrow 0$.

If instead the control time $t^*$ is fixed, in view of Eq.~\eqref{eq:min_time} the minimum perturbation strength for ${\cal C}_\alpha(t^*)$ to be sufficiently decayed is roughly
\begin{equation}
\label{eq:fgreps}
    \epsilon^* \sim  \sqrt{\frac{\hbar\ln N}{2\pi v_{\rm rms}^2\rho_g(E)t^*}}\ .
\end{equation}
Semiclassical analysis~\cite{Cerruti02, Cerruti03} generates an alternative expression to Eq.~\eqref{eq:fgreps} within the Fermi Golden rule regime:
\begin{equation}
    \label{eq:fidpred}
    \epsilon^* \sim \hbar\sqrt{\frac{\ln{N}}{2K(E)t^*}}\ ,
\end{equation}
with $K(E)$ a classical diffusion constant.  Increasing $\epsilon^*$ up to the point that the fidelity saturates shortens $t^*$ to $\tau$.  This occurs for
\begin{equation}
    \label{eq:fidpred2}
    \epsilon^* \sim \hbar\sqrt{\frac{h_{\rm KS}}{2K(E)}}\ ,
\end{equation}
where everything inside the square root is purely classical, and the strength of the perturbation needed to control the quantum dynamics scales proportionally to $\hbar$ and thus is vanishing in the $\hbar\rightarrow 0$ limit.  For this relation to hold, the control time must be chosen to be approximately the logtime, Eq.~\eqref{eq:logt}.

Up to this point in the discussion, the relations involving control time and perturbation strength scales emerge from the analysis of Peres' quantum fidelity decay for quantum chaotic systems.  However, there is a missing critical ingredient, which is ``can quantum chaos actually provide a resource for control and if so in what way?''. To address this, the arguments below involving quantum ergodicity must be taken into account.

%%%%%%%%%%%%%%%%%%%%%%%%%%%%%%%%%%%%%%%%%%%%%%%%%%%%%%%%%

\subsection{Mimicking random matrix ensembles with weak perturbations}
\label{sec:mimic}

In contrast to classical ergodicity, quantum ergodicity is often understood to mean that in the $\hbar \rightarrow 0$ limit almost all the individual eigenstate densities converge to uniform coverage of the available phase space (coarse grained over $h^D$ volumes), i.e.~the theorem of Shnirelman, Colin de Verdi\`ere, and Zelditch~\cite{Shnirelman74, Colindeverdiere85, Zelditch87}, the random wave hypothesis, and it's generalization to a microcanonical density in a Wigner representation~\cite{Berry77, Voros79}; see also~\cite{Stechel84, Peres84}.  There may be exceptional cases, such as quantum scarred eigenstates~\cite{Heller84}, but they are of vanishing measure.  The implication is that the expectation values of smooth operators over individual eigenstates generally converge to their microcanonical averages as $\hbar \to 0$.  Similarly, the intensity average of any initially localized state after a sufficiently long propagation time would also behave similarly to the individual eigenstates~\cite{Hutchinson81}.  Application to many-body systems leads to the eigenstate thermalization hypothesis~\cite{Srednicki94, Deutsch91}.

In the context of random matrix theory (RMT), which is generally used to identify whether a specific system is quantum chaotic~\cite{Bohigas84, Brody81}, especially for those systems without classical analogs, ergodicity takes on a somewhat related, but different meaning.  To say that the RMT ensembles of Wigner-Dyson~\cite{Wigner55, Dyson62a, Dyson62e} are strongly ergodic means that statistical measures calculated across the spectrum of a single member of the ensemble converge to the corresponding ensemble averages as the dimensionality of the system $N\rightarrow \infty$~\cite{Pandey79}.

There is an alternative ergodic concept of greater interest for purposes of control, which is straightforward to see from the joint probability densities for the classic Gaussian ensembles, i.e.
\begin{equation}
\label{eq:jpd0}
\rho(H) {\rm d}H = c_\beta \exp\left[-\frac{N\beta}{4}Tr\left(H^2\right)\right] {\rm d}H \ ,
\end{equation}
normalized so that the semicircular level density is radius two, and $\beta = 1, 2, 4$ for the orthogonal, unitary, and symplectic ensembles, respectively.  Using the diagonalizing similarity transformation
\begin{equation}
\Lambda = U_\beta H U_\beta^{-1} \ ,
\end{equation}
to change variables in the joint probability density generates
\begin{equation}
\label{eq:jpd}
\rho(H)  {\rm d} H \! \propto \!\! \prod_{j > k =1}^N\left| E_j-E_k\right|^\beta {\rm e}^{-\frac{N\beta}{4}\sum_{j=1}^N E_j^2}  {\rm d}\Lambda \ {\rm d}U_\beta \ ,
\end{equation}
where ${\rm d}U_\beta$ is the appropriate Haar measure (i.e.~orthogonal, unitary, symplectic), and the $\beta$ power of the Vandermonde determinant is essentially responsible for the properties of level repulsion and long range spectral rigidity~\cite{Dyson72, Pandey81, Kamien88}.

The key points are that, probabilistically speaking across the ensemble, every eigenbasis is equally likely, each eigenbasis is completely independent of the spectrum, and every non-degenerate $N$-dimensional spectrum has some weight in the ensemble, which can be viewed as a kind of ergodicity, and a resource for quantum control.  Imagine desiring to evolve an arbitrary initial state $\ket{\alpha_{\rm i}}$ into an arbitrary final state $\ket{\alpha_{\rm f}}$ (orthogonal to $\ket{\alpha_{\rm i}}$) at time $t^*$.  There is always a nonvanishing measure of RMT ensemble members that accomplish this endeavor.  For a simple example, there would be ensemble members with a pair of eigenvectors given by $\left(\ket{\alpha_{\rm i}} \pm \ket{\alpha_{\rm f}}\right)/\sqrt{2}$ possessing whatever energy separation one wished.  The shorter the desired $t^*$, the greater the energy spacing required.  This is but a single possibility from a myriad of them.  The essential difficulty of controlling the dynamics is identifying the useful Hamiltonian members of the ensemble, which after the similarity transformations lead to spectra and eigenbases suitable for the desired outcome.   Although they are numerous, they are also vanishingly rare within the ensemble (exponentially so) as $N\rightarrow \infty$.

Taken seriously, the association of RMT to strongly, fully chaotic dynamical systems implies that an ensemble of very weak perturbations should generate an ensemble behaving very much like the random matrix ensembles. Indeed, in 1983 Pechukas~\cite{Pechukas83} argued that a system with a parametrically varying Hamiltonian would generate a Vandermonde determinant-like joint probability locally in the spectrum, and in 1988~\cite{Wilkinson88}, Wilkinson imagined a dynamical ensemble due to variations of semiclassical quantizations.  Here it turns out to be sufficient to invoke weak random perturbations in order to generate a dynamical system ensemble.

For any true physical system dynamics, continuity requires non-zero time for information about the initial state to be almost entirely lost.
If there is a classical analog, then there is an Ehrenfest time  scale $\tau_{\rm E}$ \cite{Ehrenfest27}, which is roughly the logtime for chaotic systems, i.e.  $\tau_{\rm E} \approx \tau$. However, the Wigner-Dyson RMT ensembles, due to their invariance under orthogonal, unitary, or symplectic transformations, scramble any information about the initial state almost immediately on a time scale $\tau_{\rm RMT}$, approximately given by
\begin{equation}
\frac{\tau_{\rm RMT}}{\tau_{\rm H}} \approx \frac{1}{N} \approx 0 \ ,
\end{equation}
relative to the Heisenberg time $\tau_{\rm H}$; see Appendix~\ref{sec:app}. 
To have a proper non-vanishing scrambling time, there must exist weak matrix element correlations that enforce the continuity of the dynamics. The Wigner-Dyson RMT ensembles with a complete absence of matrix element correlations do not possess a true logtime $\tau>0$. 

Thus, a chaotic Hamiltonian with a weak random perturbation cannot create an ensemble that is strictly equivalent to Eq.~\eqref{eq:jpd}, as the matrix element correlations required for the existence of a logtime do exist.  Nevertheless, we argue that it comes close enough for control purposes.  First point, the level velocities associated with $\epsilon \widehat H_c$ are of order $v_{\text{rms}}$, which in terms of $\hbar$ is of $O(1)$ whereas the density of states is of $O(\hbar^{-D})$.  Thus, in changing $\epsilon$ from $0$ up to $\epsilon^*$, the system would go through innumerable avoided crossings, thus scrambling the eigenstates from the initial ones of $\widehat H_0$, which themselves appeared as though they were random to within the microcanonical constraints~\cite{Berry77, Voros79}. This leads approximately to the desired Haar measure behavior of the perturbed eigenstates. Secondly, those same avoided crossings would tend to decorrelate the association of $\widehat H_0$'s eigenstates and eigenvalues with the association found between $\widehat H_0 + \epsilon^*\widehat H_c$'s eigenstates and eigenvalues. 

Finally, it is possible to make a minor modification of Pechukas argument~\cite{Pechukas83} to account for the distinction between an ensemble of very weak random perturbations as opposed to a single parametrically varying Hamiltonian. Again considering the innumerable avoided crossings between the two systems, $\widehat H_0$ and $\widehat H_0 + \epsilon^*\widehat H_c$, the local correlations between the two spectra would be vanishingly small, and would become two independent realizations of spectra possessing a statistics incorporating the Vandermonde determinant.  The value of $\epsilon^*$ though vanishing in the $\hbar\rightarrow 0$ limit, would be large enough that any ensemble of independent $\widehat H_c$ operators would generate an ensemble extremely similar to Eq.~\eqref{eq:jpd} with its essential properties.  Note that in Eq.~\eqref{eq:jpd} the Gaussian factor is irrelevant for the eigenvalues locally in the spectrum as it just constrains the overall level density~\cite{Kamien88}.  Since every spectrum and eigenbasis occurs there, one would expect an ensemble of different $\widehat H_c$ operators to be sufficiently close to every eigenbasis and spectrum as to be a resource for quantum control.  In particular, even the presence of those exponentially rare ensemble members that are needed for control would be represented.

In the context of many-body systems, the more physically motivated embedded random matrix ensembles in which the body rank of the operators is restricted~\cite{French71, Bohigas71, Kota14}, say for example including just one- and two-body operators, it is not possible to make such a strong blanket statement about the existence of all possibilities as expressed in Eq.~\eqref{eq:jpd}.  These ensembles, unlike the Wigner-Dyson ensembles, have far fewer defining matrix element parameters than eigenvalue and eigenvector components.  On the other hand, such restricted systems with just two-body operators that possess a classical limit are generally quite capable of developing strongly chaotic dynamics with positive Lyapunov exponents.  There's no classical distinction.  Furthermore, the arguments about multiple avoided crossings and pseudo-random matrix elements between unperturbed eigenstates belonging to a local energy domain not possessing selection rules and thus connecting all states remain.  Thus, focusing on controlled dynamics that only rely on a local-in-energy part of the spectrum and associated eigenfunctions, it seems plausible that quantum chaos provides a similar resource for many-body quantum systems as for simple chaotic systems involving single particles or idealized maps.\\ 

%%%%%%%%%%%%%%%%%%%%%%%%%%%%%%%%%%%%%%%%%%%%%%%%%%%%%%%%%%

\section{Implementation with the quantum kicked rotor}
\label{sec:kr}

The key principles outlined above are showcased for the paradigmatic quantum kicked rotor in strongly chaotic and integrable regimes.  Over several decades, the model has been used to illustrate a large number of dynamical effects such as the absence of diffusion~\cite{Shepelyansky83} and the Lloyd model of Anderson localization~\cite{Fishman82}.  It has the advantage of simplicity and a tunable parameter altering the dynamics from integrable to fully chaotic~\cite{Chirikov79}.  There have been many realizations with ultracold atom experiments for reasons as diverse as exploring diffusion, decoherence, localization, the metal-insulator transition, forward and backscattering, as well as the quantum boomerang effect~\cite{Moore95, Ammann98, Oberthaler99, Chabe08, Manai15, Lemarie17, Sajjad22, Cao22}.

%%%%%%%%%%%%%%%%%%%%%%%%%%%%%%%%%%%%%%%%%%%%%%%%%%%%%%%%%%
\subsection{Hamiltonian and optimal control approach}
\label{sec:oc}

The unperturbed Hamiltonian is given by
\begin{equation}
    \label{krg}
    \widehat H_0(q,p) = \frac{\widehat p^2}{2} - \frac{K}{4\pi^2}\cos (2\pi \widehat q) \sum_{n=-\infty}^\infty \delta(t-n) \ .
\end{equation}
For kicking strengths $K$ greater than roughly $2\pi$ the analog classical system's dynamics are strongly chaotic~\cite{comment}, whereas for $K=0$, the dynamics are integrable. In the chaotic limit the Lyapunov exponent is known to be $\lambda =h_{\rm KS} \approx \ln(K/2)$~\cite{Chirikov79, Tomsovic07}.  The Floquet operator of the unitary quantum dynamics is given by
\begin{equation}
    \widehat{U}_0 = \exp\left( \displaystyle\frac{-i \widehat{p}^2}{2\hbar} \right) \; \exp\left[ \displaystyle\frac{iK}{4\pi^2\hbar}  \cos 2\pi\widehat{q} \right]\ . 
\label{eq.4a3}
\end{equation}
One way to envision the creation of a dynamical ensemble is by applying a very weak perturbation to the system in the form of spatial disorder
\begin{equation}
\label{eq:perturbation_potential}
V_{\vec{\epsilon}}(\widehat{q}) = \sum_{k=1}^N \epsilon_k\cos (2\pi k\widehat{q})\ ,
\end{equation}
where each ensemble member is characterized by a vector $\vec{\epsilon} = (\epsilon_1, ...,\epsilon_N)$ determining the strength of the disorder harmonics. The corresponding perturbed quantum dynamics is then just given by $\widehat{U} = \widehat{U}_0\widehat{U}_{\vec{\epsilon}}$ with
\begin{equation}
\label{eq:perturbed_floquet}
   \widehat{U}_{\vec{\epsilon}} = \exp\left[ \displaystyle\frac{i}{\hbar}  \sum_{k=1}^N \epsilon_k\cos (2\pi k\widehat{q}) \right] \ .
\end{equation} 
Since the system is quantized on the unit torus, the maximum number of harmonics is $N$. To have the $3N$ parameters to control the full eigenbasis and the spectrum, it is possible to introduce two weak intermediate kicks, similarly to the case constructed in Ref.~\cite{Tomsovic23b}.

%%%%%%%%%%%%%%%%%%%%%%%%%%%%%%%%%%%%%%%%%%%%%%%%%%%%%%%%
\begin{figure}
    \centering
    \includegraphics[width=1\linewidth]{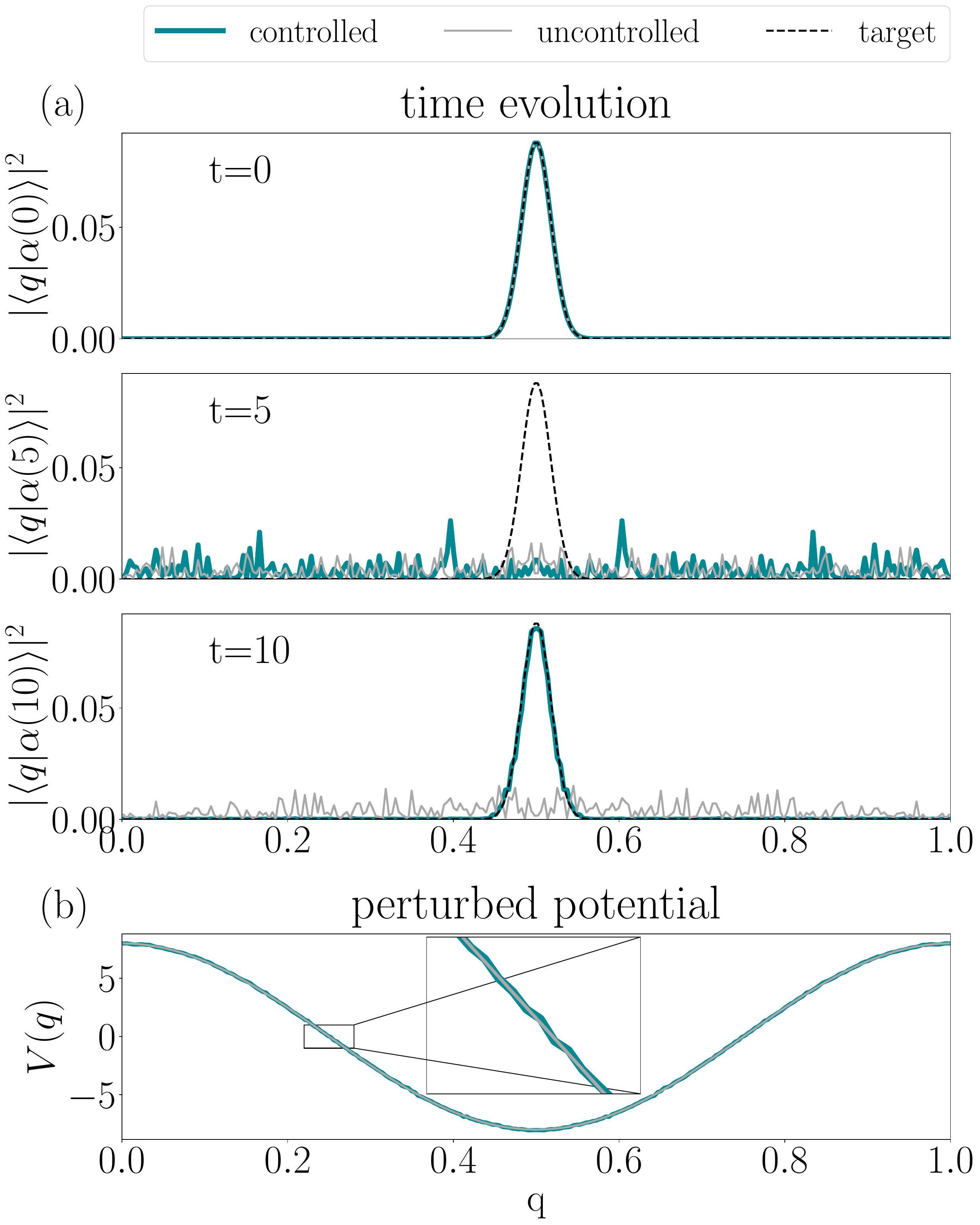}
    \caption{A revival of an initially localized state in a quantum kicked rotor ($K=8, N=256$). The state fully disperses in both controlled and uncontrolled time evolution seen in the middle panel of (a). The controlled state however reconstructs at $t=t^*=10$ with fidelity $\mathcal{F}=0.99$ (lower panel of (a)), whereas the uncontrolled state does not. The revival is compelled by the very weak ($(V_{\vec{\epsilon}})_{\rm rms}= 0.049$), barely visible, finely tuned spatial disorder displayed in (b).}
    \label{fig:rev_timeEvo}
\end{figure}
%%%%%%%%%%%%%%%%%%%%%%%%%%%%%%%%%%%%%%%%%%%%%%%%%%%%%%%%%
\begin{figure}
    \centering
    \includegraphics[width=\linewidth]{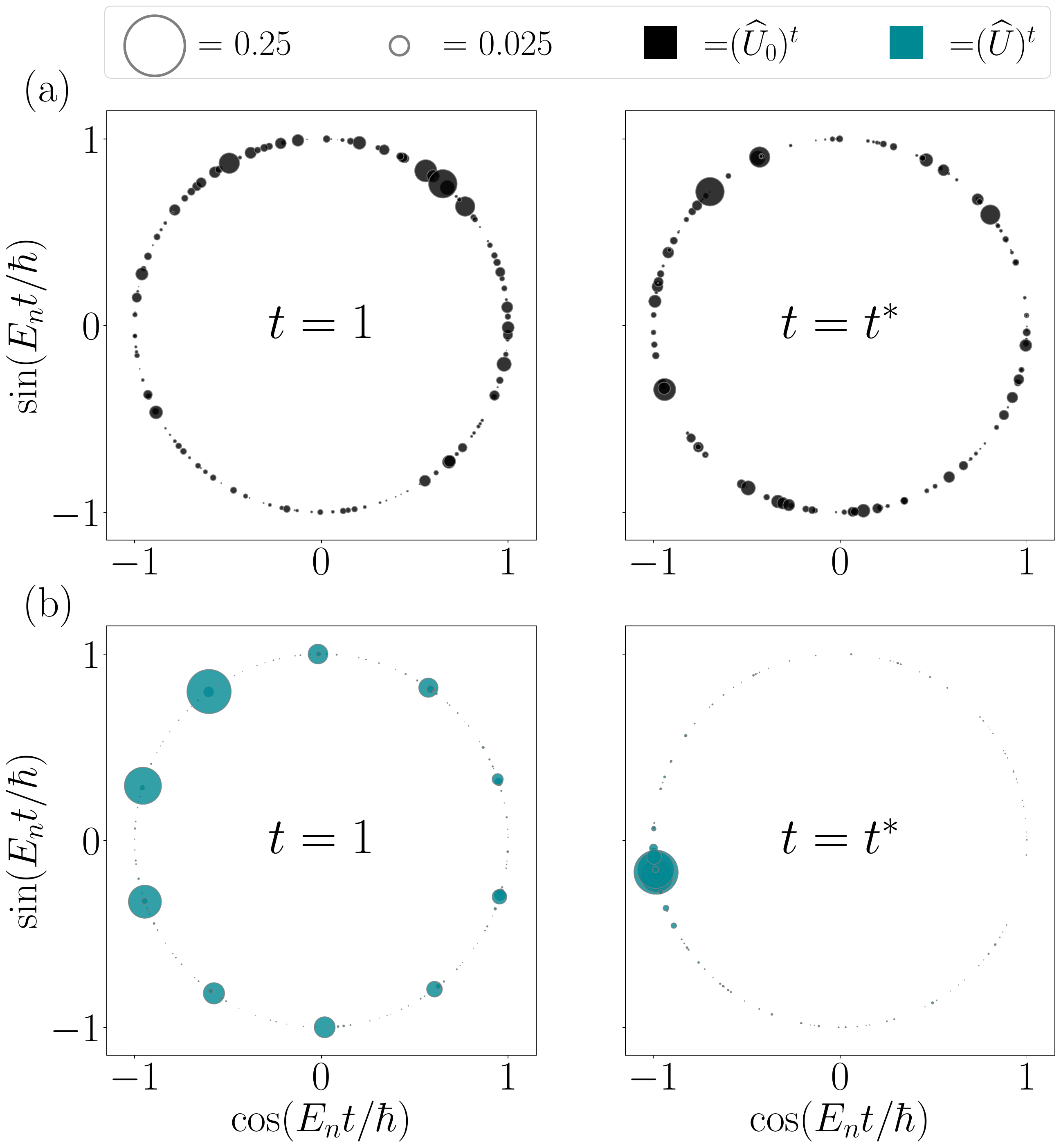}
    \caption{Quasienergy distribution of the Floquet operator $\widehat{U}^{t}$. In the uncontrolled case (a), the phases of the quasienergies and initial state intensities $|c_n|^2$ (indicated by the radius of the dots) distribute without any regular structure on the unit circle.  However, by weakly perturbing the quantum kicked rotor with a fine tuned potential (b), the phases are adjusted in just the right way, such that those corresponding to the largest intensities all line up at $t=t^*=10$. The resulting constructive interference then lead to a revival of the initial state.}
    \label{fig:rev_intensities}
\end{figure}
%%%%%%%%%%%%%%%%%%%%%%%%%%%%%%%%%%%%%%%%%%%%%%%%%%%%%%%%%%%

Suitable ensemble members can be identified using quantum optimal control~\cite{Werschnik07, Koch22, Ansel24} by minimizing the cost function 
    \begin{equation}
        S[\vec{\epsilon}] = 1-|\langle\alpha_{\text{f}}|\widehat{U}(t^*)|\alpha_{\text{i}}\rangle|^2 + \mu |\vec{\epsilon}|^2 \ ,
    \label{eq:cost_function}
    \end{equation}
where the second term is the target fidelity  $\mathcal{F}$ with $|\alpha_{\text{i}}\rangle$ being the initial state and $|\alpha_{\text{f}}\rangle$ being the target state.  The hyperparameter $\mu$ is set sufficiently large so as to penalize perturbation intensities greater than necessary in order to encourage the control procedure to converge toward the minimal perturbation strength, $\left(V_{\vec{\epsilon}}\right)_{\rm rms}$.  This permits the scaling argument of Eq.~\eqref{eq:fidpred} (or Eq.~\eqref{eq:fidpred2}) to be checked properly ahead.

In practice, Eq.~\eqref{eq:cost_function} is minimized numerically using a deterministic, gradient-free optimization algorithm (here using Powell's method). Despite its simplicity and  the large parameter space (up to $3N$), the minimization  reliably converges without fine tuning~$\mu$.  The target fidelity optimization procedures are run several times with different random initial seeds.  Each run generates a unique vector $\vec{\epsilon}$ that differs significantly in detail, though not scale, from those acquired with different initial seeds. Gratifyingly and as optimistically anticipated, the run-to-run variance of the target fidelity (between the initial seeds) remains negligibly small indicating that $\left( V_{\vec{\epsilon}}\right)_{\rm rms}$ is indeed converged sufficiently close to its minimally required scale.  The variations in the results are consistent with the dynamical ensemble property (and random matrix ensemble) that there exists an enormous number of ways in which the unitary evolution can be directed towards the target state; see the arguments given in Sec.~\ref{sec:mimic} after Eq.~\eqref{eq:jpd}.

Naturally, to apply optimal control with ever increasing system sizes eventually becomes challenging, especially if having many-body systems and their exponentially growing Hilbert spaces in mind. To this end it would be necessary to find algorithms that identify the suitable ensemble in a more efficient way with respect to the scaling with the Hilbert space dimension.

%%%%%%%%%%%%%%%%%%%%%%%%%%%%%%%%%%%%%%%%%%%%%%%%%%%%%%%%%

\subsection{Control in the chaotic regime: reviving, cat-like, and random states}
\label{sec:cc}

\textit{Reviving state:}
A special control example is the revival of an initially localized state $|\alpha\rangle$ at an arbitrarily chosen time $t^*=10$ as shown in Fig.~\ref{fig:rev_timeEvo}. For this example, only the $N$ eigenenergies need to be controlled (no intermediate kicks), because every system whose spectrum fulfills the relation $E_n t^* = 2\pi \hbar  m_n +\phi$, creates a revival, since at this time
\begin{equation}
\label{eq:rev_condition}
    |\langle\alpha|\widehat{U}(t^*)|\alpha\rangle| = \left|\sum_n |c_n|^2 e^{-\frac{i}{\hbar}E_n t^*}\right| = \sum_n |c_n|^2 = 1 \ .
\end{equation}
This means that it is sufficient to introduce spatial disorder to the main kick using Eq.~\eqref{eq:perturbation_potential}. The necessary root mean square value of the perturbation $(V_{\vec{\epsilon}})_{\rm rms}$ is less then one percent of the kicking strength $K$, such that the perturbed potential is almost imperceptibly different from the unperturbed one; see Fig.~\ref{fig:rev_timeEvo}.  As expected for strongly chaotic dynamics, the state quickly spreads out over the whole configuration space.  At an intermediate time there is no way to identify which state is controlled and which is uncontrolled. However at $t=t^*$ the controlled state fully reconstructs.
As Fig.~\ref{fig:rev_intensities} reveals, the finely tuned disorder lines up the phases of the largest intensities $|c_n|^2$ at $t^*$, forcing the constructive interference of Eq.~\eqref{eq:rev_condition}.

%%%%%%%%%%%%%%%%%%%%%%%%%%%%%%%%%%%%%%%%%%%%%%%%%%%%%%%%%%
\begin{figure}
    \centering
\includegraphics[width=1\linewidth]{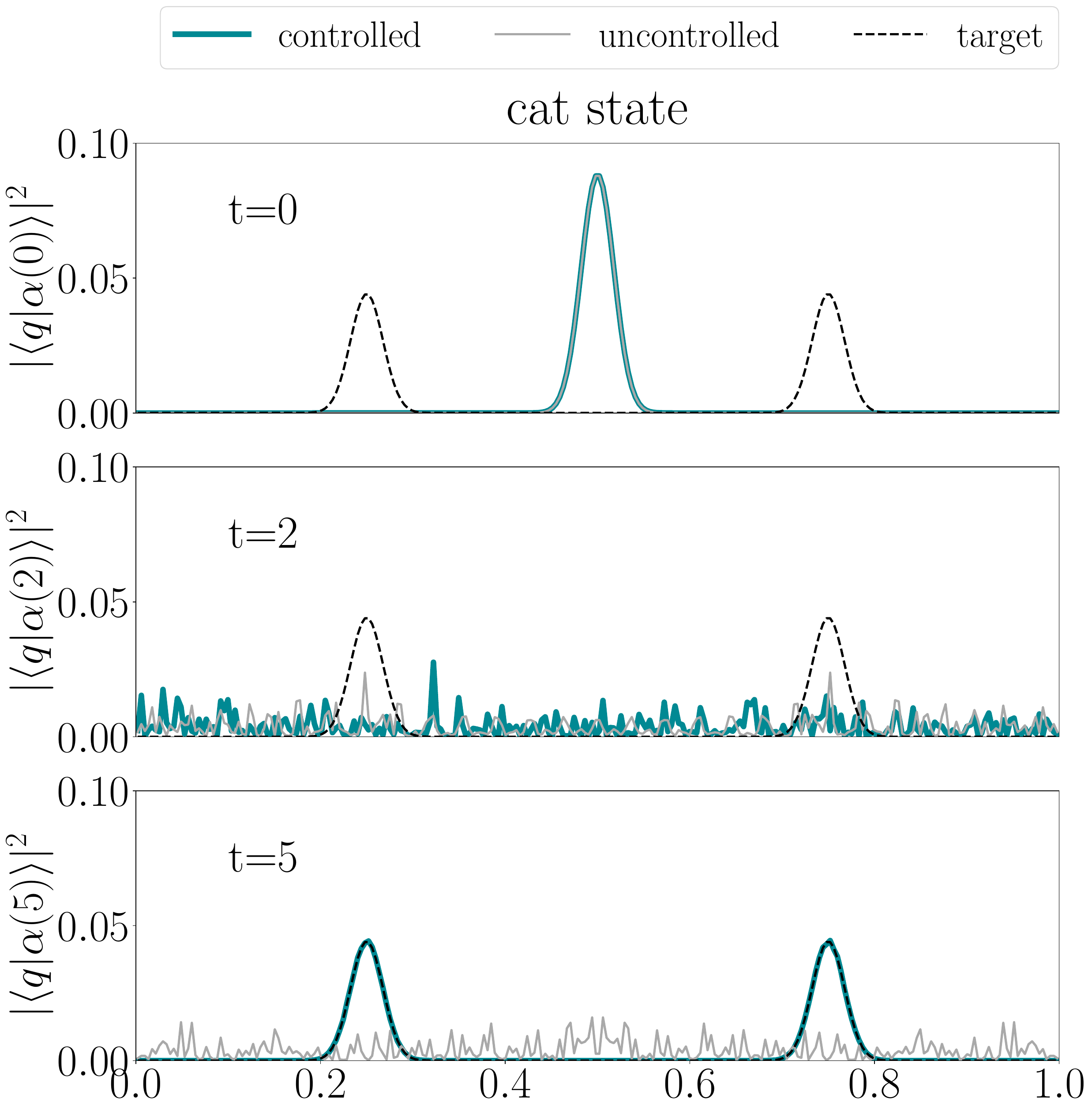}
    \caption{An initially localized state steered toward a cat-like state ($K=8, N=256, t^*=5$).  Similarly to the revival case, Fig.~\ref{fig:rev_timeEvo}, the initial state fully disperses before emerging as a cat-like state at the control time.  For full state controllability, the main kick is weakly perturbed and two additional weak kicks are inserted in between main kicks.  The finely tuned weak perturbation has a $(V_{\vec{\epsilon}})_{\rm rms}=0.0115$, and the initial state is directed almost perfectly ($\mathcal{F}>0.999$) toward the desired cat state.  The uncontrolled evolution remains random looking at the control time.}
    \label{fig:cat_evo}
\end{figure}
%%%%%%%%%%%%%%%%%%%%%%%%%%%%%%%%%%%%%%%%%%%%%%%%%%%%%%%%%%%%%%%%%%

\textit{Cat-like target state:}
Since the ergodic, RMT-like structure of the dynamical ensemble implies full state controllability, any arbitrary target can be prepared in the quantum kicked rotor by just weakly perturbing the potential (and adding two intermediate weak kicks to have sufficiently many free parameters). This includes interesting cases like well-defined superpositions as shown in Fig.~\ref{fig:cat_evo}, where instead of reviving, the initially localized state is steered toward a cat state, as schematically shown in Fig.~\ref{fig:schematic}. Again the fine-tuned perturbation is so weak that it is not worth displaying, since the difference is not really visible just as in Fig.~\ref{fig:rev_timeEvo}b).

%%%%%%%%%%%%%%%%%%%%%%%%%%%%%%%%%%%%%%%%%%%%%%%%%%%%%%%%%%
\begin{figure}
    \centering
\includegraphics[width=1\linewidth]{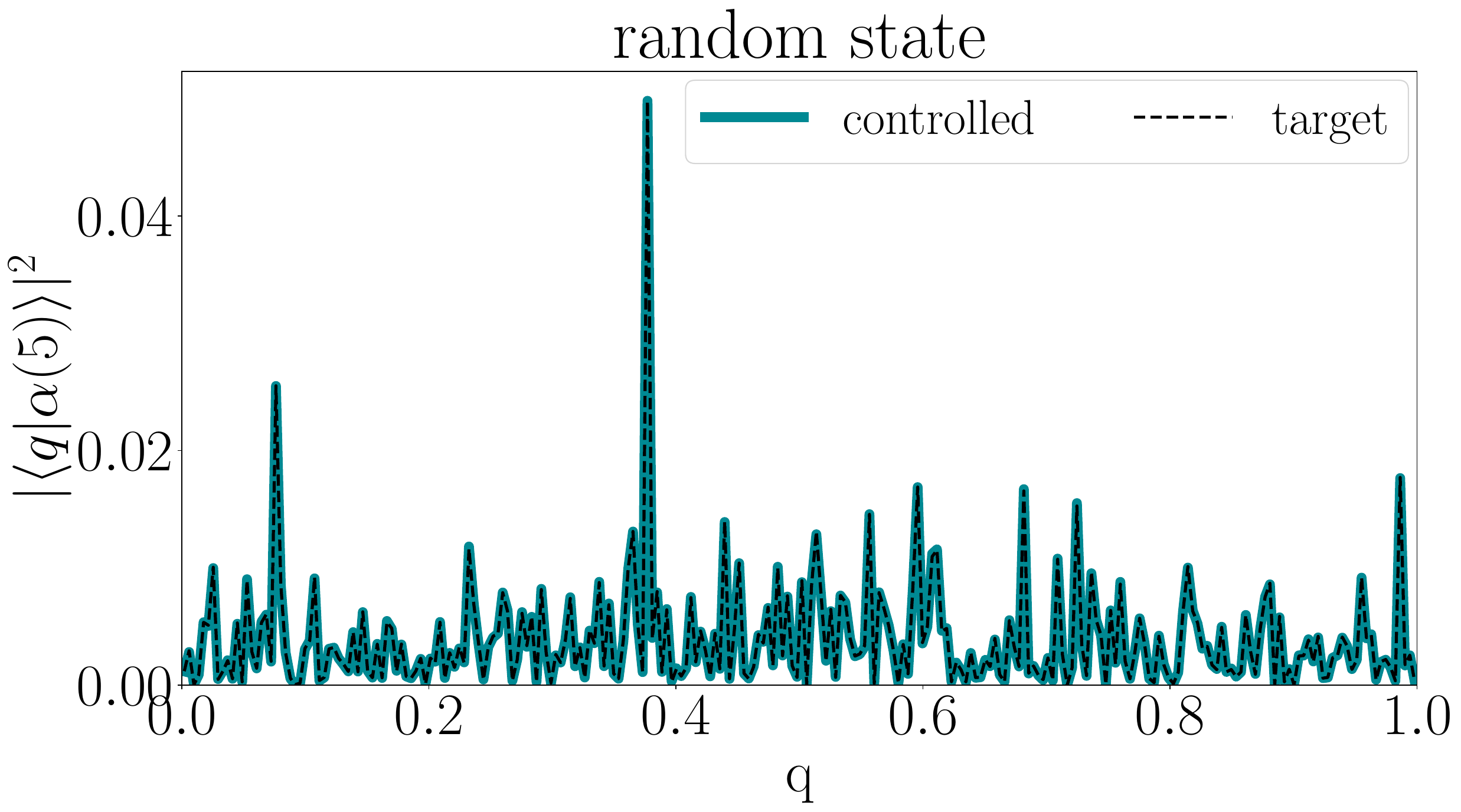}
    \caption{An initially random state is steered toward an orthogonal random state ($K=8, N=256, t^*=5$).  As it evolves, the initial state goes through a series of random-looking states settling on the target state at the control time.  For full state controllability, the main kick is weakly perturbed and two additional weak kicks are inserted in between main kicks.  The finely tuned weak perturbation has a $(V_{\vec{\epsilon}})_{\rm rms}=0.0094$, and the initial state is directed almost perfectly ($\mathcal{F}=0.999$) toward the desired random state.}
    \label{fig:rand_evo}
\end{figure}
%%%%%%%%%%%%%%%%%%%%%%%%%%%%%%%%%%%%%%%%%%%%%%%%%%%%%%%%%

\textit{Random initial and target states:}
As expected the initial and final state do not need to be localized.  In fact, random amplitudes for the initial and target state can be selected as shown in Fig.~\ref{fig:rand_evo}. Neither the initial random state, the intermediate evolved state, nor the perturbed potential are illustrated there since they behave exactly as in the earlier examples.  For the three cases presented here, i.e.~Fig.~\ref{fig:rev_timeEvo} to \ref{fig:rand_evo}, the achieved target fidelity is so nearly perfect ($\mathcal{F} > 0.99$) that the difference in the overlap between the controlled time evolved state and the target state is barely visible.

\begin{figure}
    \centering
    \includegraphics[width=1\linewidth]{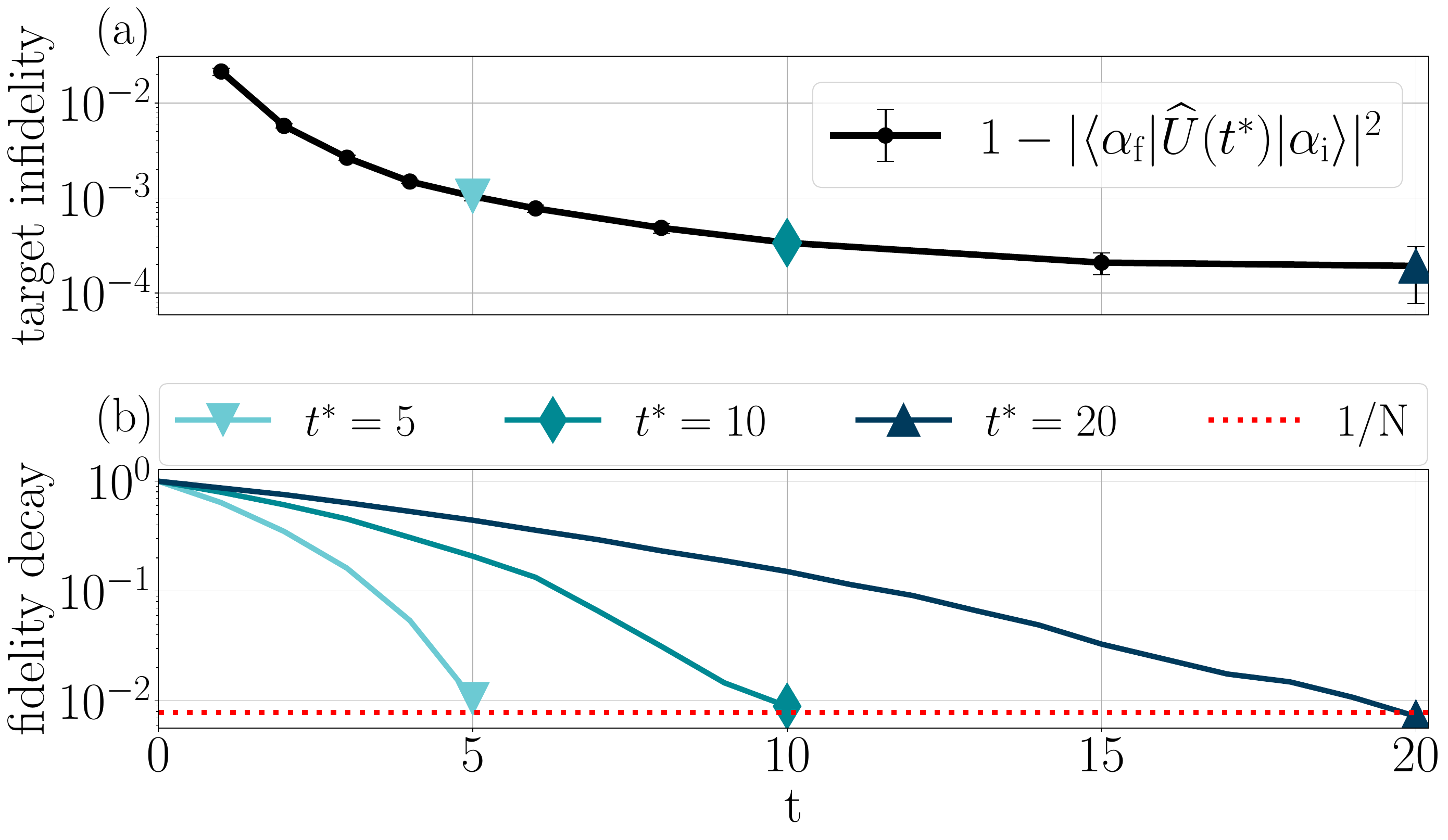}
    \caption{Full state controllability in the strongly chaotic quantum kicked rotor ($N=128, K=8, \mu = 25$). A sample of randomly picked, orthogonal initial and target states can be controlled by perturbing the main kick and adding two weak intermediate kicks.  After a short time the target fidelity (a) saturates to one and its variance over the sample (error bars) disappears, indicating perfect controllability independent of initial and target state.  When looking at the fidelity decay (b) for a perturbation $\vec{\epsilon}$ corresponding to a specific control case,  $\mathcal{C}_{\alpha}(t)$ decays to $1/N$ by the control time as shown for $t^*=5,10,20$.}
    \label{fig:fidelity_vs_time}
\end{figure}

\begin{figure}
    \centering
\includegraphics[width=1\linewidth]{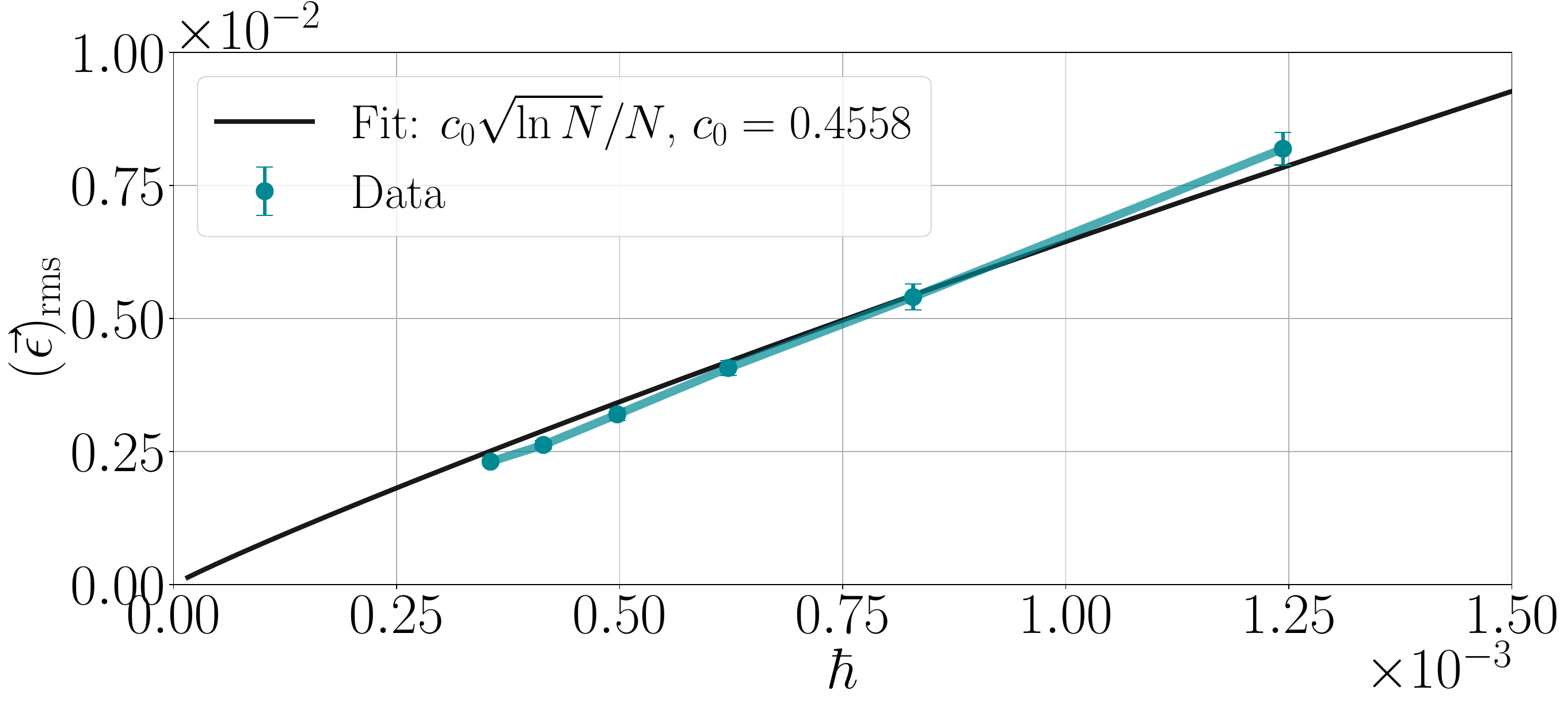}
    \caption{Strength of the perturbation in the $\hbar\to 0$  limit (large $N$). The root mean square of the perturbation $\vec{\epsilon}$ needed to control any initial state into any target state ($t^*=5$) in the chaotic regime of the quantum kicked rotor ($K=8$) decreases approximately linearly with shrinking $\hbar=1/(2\pi N)$ as predicted by fidelity decay arguments.}
    \label{fig:rms_amplitudes}
\end{figure}

That the quantum chaotic universal behavior makes the system fully controllable through only weak perturbations is further supported by Fig.~\ref{fig:fidelity_vs_time}. The target fidelity (individually optimized and averaged over a sample of randomly picked, orthogonal initial and target states), saturates after a very short time. This is only possible because the optimized disorder, $\vec{\epsilon}$, which maximizes the target fidelity, is enough for ${\cal C}_\alpha(t)$ to decay to $1/N$ by the control time $t^*$, which is one of the main assertions in the above discussion. Due to the white noise structure of Eq.~\eqref{eq:perturbation_potential}, the disorder potential does not lend itself to semiclassical treatment without accounting for the diffraction due to the high harmonic components. Nevertheless, the perturbation strength as measured by the root mean square of $\vec{\epsilon}$ still vanishes in the limit $\hbar\to 0$ as predicted by Eq.~\eqref{eq:fidpred}; see Fig.~\ref{fig:rms_amplitudes}. If the form of the perturbation applied to the system allows for semiclassical analysis, one should be able to predict the slope of Fig.~\ref{fig:rms_amplitudes}, as well as the minimum control time as measured by the logtime analytically based on Eq.~\eqref{eq:logt} and Eq.~\eqref{eq:fidpred2} respectively.

%%%%%%%%%%%%%%%%%%%%%%%%%%%%%%%%%%%%%%%%%%%%%%%%%%%%%%%5
%%%%%%%%%%%%%%%%%%%%%%%%%%%%%%%%%%%%%%%%%%%%%%%%%%%%%
%%%%%%%%%%%%%%%%%%%%%%%%%%%%%%%%%%%%%%%%%%%%%%%%%%%%%%%%

\subsection{Control in the integrable regime}
\label{sec:int}

\begin{figure}
    \centering
    \includegraphics[width=1\linewidth]{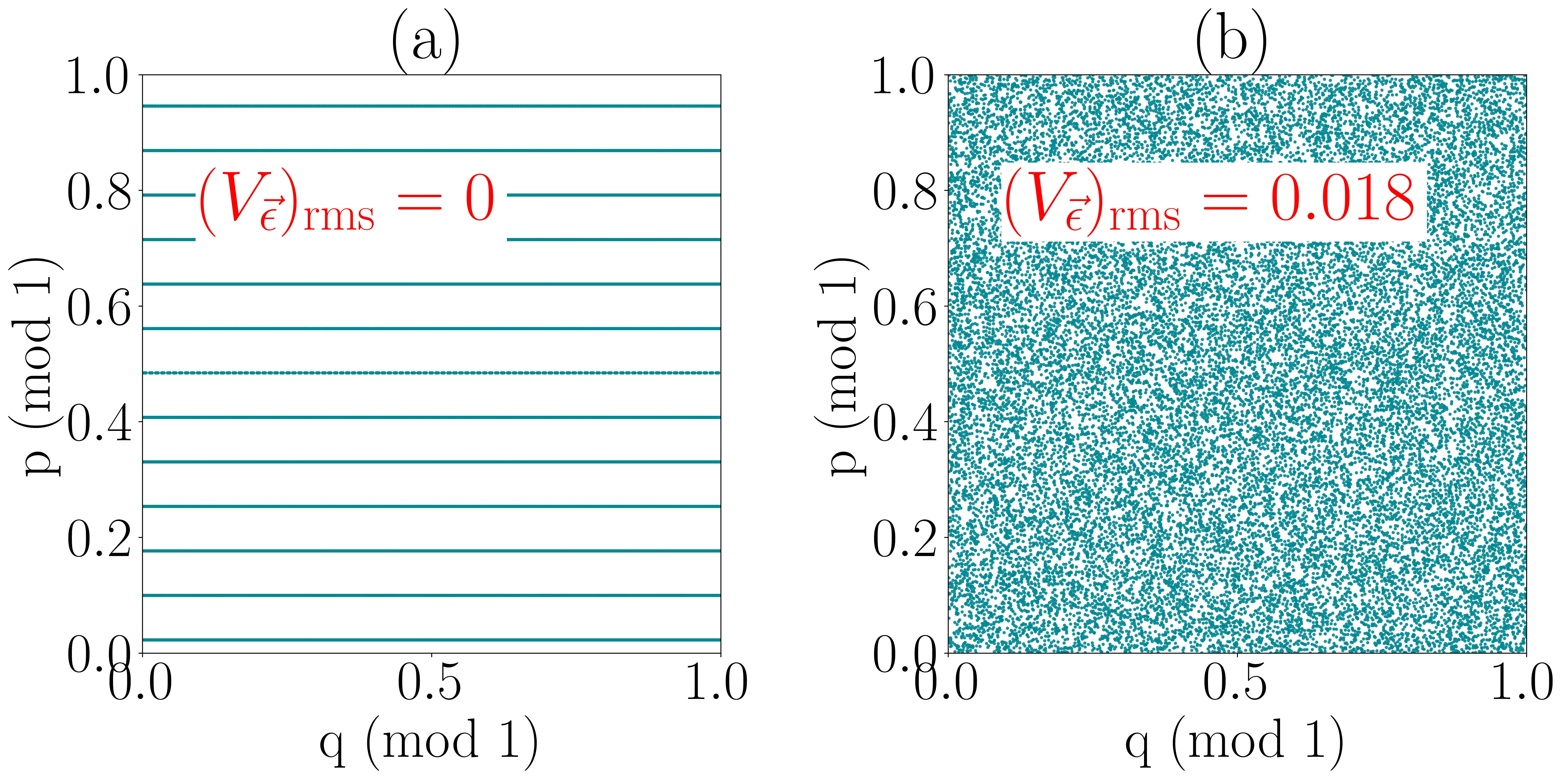}
    \caption{Integrability breaking through weak disorder. The classical kicked rotor is completely integrable for $K=0$, which is highlighted by the surface of section in panel (a); the trajectories are stable and form straight lines. Nonetheless, arbitrary state preparation is possible with high fidelity ($\mathcal{F}>0.999$) by applying only a weak disorder potential derived from minimizing Eq.~\eqref{eq:cost_function}. A perturbation of this form, even though small in intensity, is enough to alter the dynamics into a strongly chaotic regime, as shown by the chaotic sea in the surface of section in panel (b).}
    \label{fig:perturbed_poincare}
\end{figure}

\begin{figure}
    \centering
    \includegraphics[width=1\linewidth]{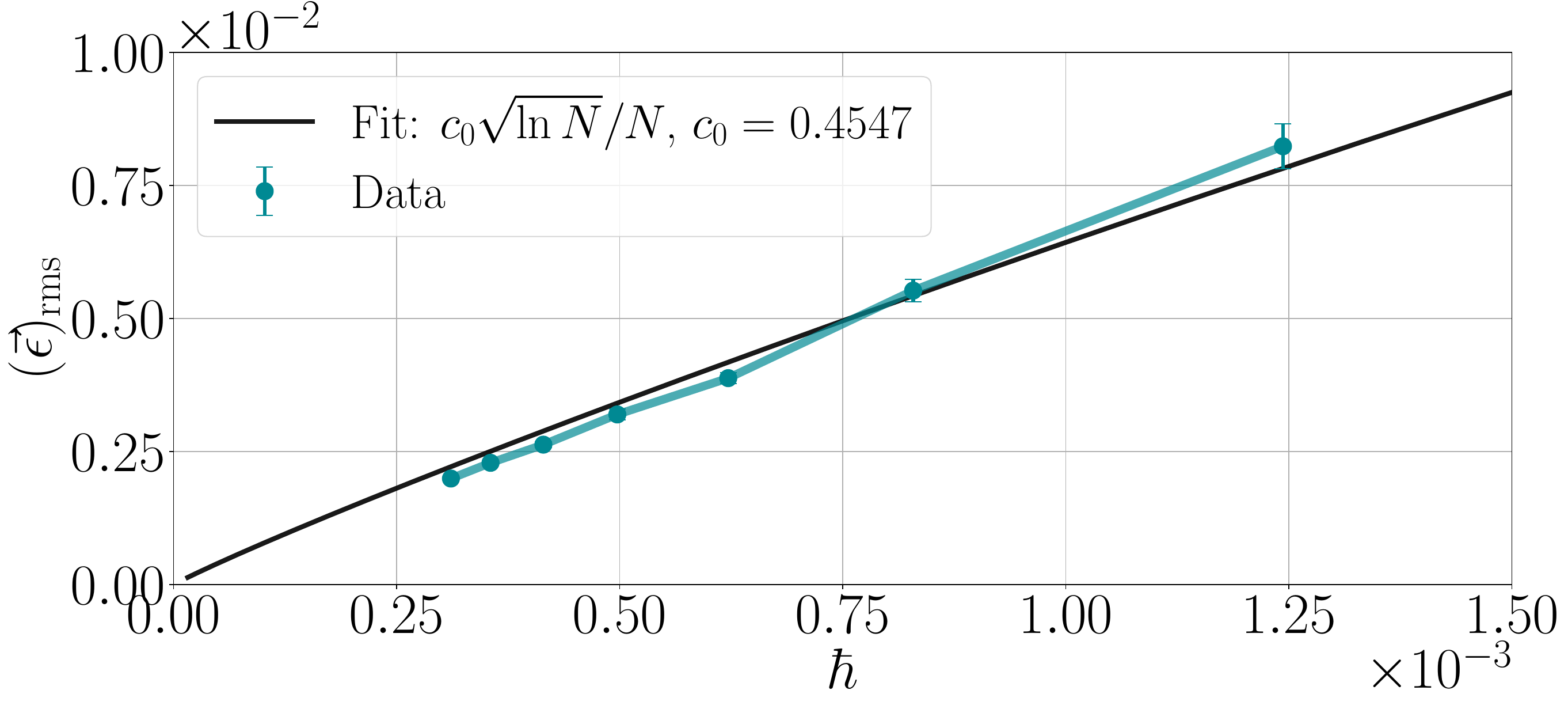}
    \caption{Strength of the perturbation in the $\hbar\to 0$  limit (large $N$).  The root mean square of the perturbation $\vec{\epsilon}$ needed to control any initial state into any target state ($t^*=5$) decreases approximately linearly with shrinking $\hbar=1/(2\pi N)$ even in the integrable regime of the quantum kicked rotor ($K=0$).}
    \label{fig:rms_amplitudes_integrable}
\end{figure}

Since neither the ergodicity nor the chaotic dynamical fidelity decay behaviors discussed can be expected to hold for integrable systems~\cite{Gorin06}, there is no equivalent argument that $\epsilon^*$ vanishes in the $\hbar \to 0$ limit. A priori, it may seem that this implies that initially integrable dynamical systems lack controllability through weak perturbations.  However, a perturbation that effectively breaks integrability would bring the system back to the above argumentation for chaotic dynamical systems.  Recall that although individual trajectories in integrable systems are stable, integrable systems are structurally unstable.  Perturbations tend to destroy integrability~\cite{Lichtenberg83}, and even small random, disorder-like perturbations can very effectively accomplish this task (chaotic systems are the opposite, trajectories are unstable, but the system dynamics is structurally stable~\cite{Cerruti02}, i.e.~perturbations tend to leave the system chaotic). Perhaps, it should not be surprising that the quantum kicked rotor exhibits full controllability even for the integrable case $K=0$. It turns out that, in fact, any potential of the form of Eq.~\eqref{eq:perturbation_potential} with $(\vec{\epsilon})_{\rm rms} > \epsilon^*$ is sufficient to alter the dynamics from integrable to strongly chaotic, see Fig.~\ref{fig:perturbed_poincare}.  This is to be expected from the nature of this disorder perturbation, which tends towards a white noise potential with $N \to \infty$, since each additional high $k^{th}$ harmonic contributes to breaking integrability quadratically with respect to $k$.  For smooth perturbations, it may not be possible to arrange controllability with $\epsilon^* \to 0$, however, for the disorder potential applied to the $K=0$ case, Fig.~\ref{fig:rms_amplitudes_integrable} shows the same behavior as for the chaotic regime.
%%%%%%%%%%%%%%%%%%%%%%%%%%%%%%%%%%%%%%%%%%%%%%%%%%%%%%%%

%%%%%%%%%%%%%%%%%%%%%%%%%%%%%%%%%%%%%%%%%%%%%%%%%%%%%%%

%%%%%%%%%%%%%%%%%%%%%%%%%%%%%%%%%%%%%%%%%%%%%%%%%%%%%%%%%
\section{Summary}
\label{sec:sum}

In conclusion, a quantum chaotic system's dynamical sensitivity to weak perturbations and its ability to mimic random matrix ensembles enables full controllability on very short time scales. Only a very weak perturbation is required with a strength vanishing more or less linearly as $\hbar \to 0$.  The quantum kicked rotor nicely illustrates the main predicted features where control worked extremely well.  There are no limitations on the choice of initial and final states so long as they respect the same symmetries and lie within the same Hilbert space. It also happened that not much difference occurred starting with the integrable kicked rotor, but closer inspection showed that the white-noise-like weak perturbation was sufficient to destroy integrability in that case.

Decades ago work began on showing that exponential instability and ergodicity enable the control of classically chaotic dynamical systems~\cite{Ott90}.  The question has remained, however, as to whether there is a parallel for quantum chaotic systems.  Since the exponential instability and ergodicity concepts do not directly translate from classical mechanics into quantum mechanics, it is first necessary to identify corresponding or analogous features.  Semiclassical theories of Peres' fidelity decay have established generic regimes that describe the quantum sensitivity to perturbations~\cite{Jalabert01, Cerruti02, Cerruti03}.  This sensitivity gives rise to decay features and scales that serve as the closest corresponding feature to the exponential instability of classically chaotic systems.  

As for ergodicity, the most useful analogous quantum feature is the ability of perturbations to quantum chaotic systems to mimic closely three features of random matrix ensembles, i.e.~within a relevant Hilbert space of some effective dimensionality, all eigenbases are equally probable, the eigenbases have no correlations with the spectrum, and finally, every spectrum has some weight in the ensemble.  Together with the sensitivity mentioned above, these features enable rapid full controllability.    Dynamical continuity prevents physical systems from having a perfect realization of these random matrix properties, but the arguments presented above suggest that for evolution beyond the so-called logtime a constructed dynamical ensemble is sufficiently close to the random matrix ideal.

Recall that the logtime is an extremely short timescale and that it grows very slowly with $\hbar\to 0$ and system size or numbers of degrees of freedom tending to infinity, and thus it implies that control can be achieved extremely quickly even in many-body systems with large numbers of degrees of freedom; see Appendix~\ref{sec:app2}.  In fact, in the context of full controllability there are similarities between the Mandelstam-Tamm bound~\cite{Mandelstam45} and the logtime.  For example, the perfect isotropy of the Wigner-Dyson ensembles guarantees that the control time and the Mandelstam-Tamm bound would be equivalent; see Appendix~\ref{sec:app}.  However, due to dynamical continuity (lack of perfect isotropy) the logtime, which occurs later than the Mandelstam-Tamm bound, is the time at which any and every target state can be reached from any and every initial state through control with a vanishing perturbation strength as $\hbar\to 0$.

There are a number of directions of possible future research.  The distinction between embedded random matrix ensembles, where the body rank of the operators are limited, and the Wigner-Dyson ensembles also suggests that there cannot be perfect isotropy for a further reason in dynamical ensembles constructed with many-body systems.  The ideas presented here would be straightforward to apply to, say, the Bose-Hubbard model or spin chains in order to see whether there are further subtleties for such systems.  It would also be interesting to investigate implications for the purposes of quantum computing, such as the control of quantum gates. There has been a great deal of interest in random unitary evolutions, e.g.~see~\cite{Emerson03, Dankert09, Mele24, Cui25, Zhou25},
in the work described above any random initial state can be rapidly evolved into any other with straightforward optimal control methods (e.g.~Fig.~\ref{fig:rand_evo}).

Another direction involves the full development of semiclassical theory for the control process.  In cases where the system has a well defined classical analog and the ensemble of perturbations involves smooth operators, first order semiclassical perturbation theory is applicable.  There the trajectories of the unperturbed system can be used to calculate the phase changes due to the perturbations.  That would generate a linear system of algebraic equations whose solutions should give an alternate route to optimizing control parameters to the usual optimal control equations.  In this way it may be possible to treat much greater parameter numbers and even identify a reduced parameter set. 

%%%%%%%%%%%%%%%%%%%%%%%%%%%%%%%%%%%%%%%%%%%%%%%%%%%%%%%%%5

\section{Acknowledgments}

We thank N.~Beato, D.~Gu\'ery-Odelin, G.~Morigi, D.~Ullmo  and J.~D.~Urbina for useful conversations.  We gratefully acknowledge support by the Deutsche Forschungsgemeinschaft (DFG, German Research Foundation), project Ri681/15-1 within the Reinhart-Koselleck Programme, and by the Vielberth Foundation for financial support. We further acknowledge support from the Oﬃce of Naval Research (ONR), grant N62909-24-1-2053, within the ONR Global Program.  MS is funded through a fellowship by the Studienstiftung des Deutschen Volkes.

%\section{Data availability}

%%%%%%%%%%%%%%%%%%%%%%%%%%%%%%%%%%%%%%%%%%%%%%%%%%%%%%%%%%%%5
\appendix

\section{Brevity of logtime}
\label{sec:app2}

A critical feature of the logtime is its brevity and how slowly it scales with $\hbar\to 0$ and system size or numbers of degrees of freedom tending to infinity. Note this implies that full controllability can theoretically be achieved extremely quickly even in many-body systems with large numbers of degrees of freedom.  To understand the logtime's scaling, insert the relationship between the effective Hilbert space dimension $N$, total phase space volume, and Planck cell volume of Eq.~\eqref{eq:Nth} into Eq.~\eqref{eq:logt}.  This generates
\begin{equation}
\label{eq:app2-1}
\tau = \frac{1}{h_{\rm KS}}\ln \frac{{\cal V}_{\rm Th}}{g h^D}
\end{equation}
Since the Kolmogorov-Sinai entropy is the sum of the positive Lyapunov exponents, $h_{\rm KS}=\sum_{j=1}^{D-1} \lambda_{j}$, and there are $D-1$ of them, for large $D$, the entropy more or less equals the number of degrees of freedom times the mean Lyapunov exponent, $\bar \lambda$, i.e.~$h_{\rm KS}=D\bar \lambda$.  Also, by defining an effective phase space volume per degree of freedom, ${\cal V}_1$, such that ${\cal V}_{\rm Th}={\cal V}_1^D$, the $D$-dependence of Eq.~\eqref{eq:app2-1} can be largely removed giving
\begin{equation}
\label{eq:app2-2}
\tau = \frac{1}{\bar \lambda}\ln \frac{{\cal V}_1}{h}
\end{equation}
($g^{1/D}\approx 1$).  To imagine an example logtime, consider an ideal gas in a cubic box whose interactions generate collisions which are almost point-like.  For a fixed momentum per degree of freedom, $\bar \lambda$ would be roughly proportional to the density since its origin is due to collisions, and ${\cal V}_1/h$ would just count the mean number of de Broglie wavelengths of a particle across the box length.  Thus, the logtime at constant pressure and temperature would roughly just be logarithmic in the de Broglie wavelength count across the box length. 

\section{Wigner-Dyson Ensemble Scrambling Time}
\label{sec:app}

The Wigner-Dyson ensembles generate an unphysical dynamics in the sense that an equivalent of the logtime (or equivalently a scrambling time scale) found in chaotic dynamical systems lacks the logarithmic growth with $N$, e.g.~Eq.~\eqref{eq:logt}.  Whereas a localized initial state in a dynamical system cannot reach every orthogonal state in the same time due to continuity of its dynamics, hence the logtime, these ensembles can reach every initially orthogonal state due to the perfectly flawless isotropy of these ensembles. The isotropy guarantees that any state can be reached in the same timescale that the autocorrelation function decays to $1/N$. Comparing the autocorrelation function decay to the Heisenberg timescale by considering their ratio, gives a dimensionless scale for the decay, which is suitable for comparison with a true dynamical system.

As mentioned earlier in the text, to within $O(1/N)$ corrections the eigenvalue density is a semicircle of radius two for the Wigner-Dyson ensembles normalized as in Eq.~\eqref{eq:jpd0}.  Therefore, the unit-normalized level density, $\rho(E)$, is given by
\begin{equation}
\label{eq:levden}
\rho(E) = \frac{1}{2\pi}\sqrt{4-E^2} \ .
\end{equation}
Since the mean eigenvalue spacing, $D(E)$, is the inverse of the density, for energies not too far from the center of the semicircle, $D(E) \approx \pi/N$.  This is used below to calculate both the Heisenberg timescale and the scrambling time.

In the evolution operator, let the timescale at which neighboring levels typically differ by a $2\pi$ phase difference be the Heisenberg time scale, $\tau_{\rm H}$.  Straightforward algebra generates the following approximation for the bulk of the eigenvalues near the center of the spectrum, 
\begin{equation}
\label{eq:tauh}
\tau_{\rm H} = \frac{2\pi \hbar}{D(E)} \approx 2\hbar N \ .
\end{equation}
As for the autocorrelation function of some specialized or localized state, $\ket{\alpha}$, no matter how it is defined, across the Wigner-Dyson ensemble it would appear as a Haar-random state relative to the eigenbasis.  Hence, the autocorrelation function, see Eq.~\eqref{eq:rev_condition}, has no correlation between its spectra and eigenstates, and thus the ensemble average between the eigenstate expansion coefficients and spectrum separates. Each coefficient averages to $1/N$, thus
\begin{align}
\label{eq:autocor2}
\overline{\braket{\alpha|\alpha(t)}} &= \frac{1}{N}\overline{\sum_{j=1}^N  \exp\left(- \frac{iE_j t}{\hbar}\right)} \approx \int_{-2}^2{\rm d}E \ \rho(E) {\rm e}^{-iEt/\hbar}\nonumber \\
&= \hbar\ \frac{J_1(2t/\hbar)}{t}\ .
\end{align}
Let the autocorrelation function decay timescale, $\tau_{\rm RMT}$, correspond roughly to the first zero of $J_1(2t/\hbar)$.  Then, 
\begin{equation}
\label{eq:rmthratio}
\frac{\tau_{\rm RMT}}{\tau_{\rm H}} = \frac{1.9\hbar}{2\hbar N} \approx \frac{1}{N}\ .
\end{equation}
Compare this result to the logtime-Heisenberg timescale ratio for the kicked rotor, which is given by
\begin{equation}
\label{eq:krratio}
\frac{\tau}{\tau_{\rm H}} = \frac{\ln N}{\lambda N}
\end{equation}
The only way to roughly equate these expressions would lead to the identification of the effective Lyapunov exponent (or Kolmogorov-Sinai entropy) of the Wigner-Dyson ensembles equaling $\ln N$ or tending toward infinite instability with $\hbar \to 0$ and/or system size.

\bibliography{classicalchaos,furtherones,general_ref,molecular,quantumchaos,rmtmodify,manybody, comment}

\end{document}